\documentclass[times, twocolumn]{aastex63}

\usepackage{xspace}
\usepackage{amsmath}
\usepackage{graphicx}
\usepackage{bm}
\usepackage{tikz}
\let\tablenum\relax
\usepackage{siunitx}
\newcommand{\TESS}{\emph{TESS}\xspace}
\newcommand{\Kepler}{\emph{Kepler}\xspace}
\newcommand{\gaia}{\emph{Gaia}\xspace}
\newcommand{\tic}{TIC-464300749\xspace}
\newcommand{\ticb}{TIC-464300749b\xspace}
\newcommand{\toi}{TOI-3362\xspace}
\newcommand{\toib}{TOI-3362b\xspace}
\newcommand{\ut}{UT}
\newcommand{\filterzs}{\ensuremath{z_\mathrm{s}}}
\newcommand{\filterrc}{\ensuremath{R_\mathrm{c}}}
\providecommand{\msun}{\ensuremath{M_\Sun}}
\providecommand{\rsun}{\ensuremath{R_\Sun}}
\providecommand{\lsun}{\ensuremath{L_\Sun}}
\providecommand{\rj}{\ensuremath{R_{\rm Jup}}}
\providecommand{\mj}{\ensuremath{M_{\rm Jup}}}

\newcommand{\chiron}{CHIRON\xspace}
\newcommand{\astep}{ASTEP\xspace}
\newcommand{\lco}{LCOGT-SSO\xspace}
\newcommand{\minerva}{Minerva-Australis\xspace}

\newcommand{\period}{18.09547$^{+0.00003}_{-0.00003}$}
\newcommand{\midt}{1529.325$^{+0.001}_{-0.001}$}
\newcommand{\mpl}{5.029$^{+0.668}_{-0.646}$}
\newcommand{\depth}{0.0041$^{+0.0002}_{-0.0002}$}
\newcommand{\rpl}{1.142$^{+0.043}_{-0.041}$}
\newcommand{\bp}{0.270$^{+0.207}_{-0.184}$}
\newcommand{\ecc}{0.815$^{+0.023}_{-0.032}$}
\newcommand{\argperi}{50.873$^{+11.085}_{-9.165}$}
\newcommand{\inc}{89.140$^{+0.584}_{-0.668}$}
\newcommand{\aplt}{0.153$^{+0.002}_{-0.003}$}
\newcommand{\afinal}{0.051$^{+0.008}_{-0.006}$}
\newcommand{\pfinal}{3.526$^{+0.823}_{-0.584}$}

\newcommand{\chironoffset}{6450$^{+36}_{-37}$}
\newcommand{\minervaoffset}{7617$^{+60}_{-62}$}
\newcommand{\chironjitter}{3$^{+31}_{-3}$}
\newcommand{\minervajitter}{5$^{+50}_{-5}$}
\newcommand{\chirontrend}{0$^{+4}_{-3}$}
\newcommand{\minervatrend}{-1$^{+7}_{-7}$}

\newcommand{\utessa}{0.18$^{+0.09}_{-0.09}$}
\newcommand{\utessb}{0.30$^{+0.10}_{-0.10}$}
\newcommand{\uRca}{0.23$^{+0.10}_{-0.10}$}
\newcommand{\uRcb}{0.39$^{+0.10}_{-0.09}$}
\newcommand{\uBa}{0.46$^{+0.10}_{-0.10}$}
\newcommand{\uBb}{0.30$^{+0.10}_{-0.10}$}
\newcommand{\uzsa}{0.13$^{+0.10}_{-0.10}$}
\newcommand{\uzsb}{0.38$^{+0.10}_{-0.10}$}

\newcommand{\tessyones}{0.71$^{+0.04}_{-0.04}$}
\newcommand{\tessyoneGPsigma}{3.83$^{+2.27}_{-1.16}$}
\newcommand{\tessyoneGPrho}{3.11$^{+2.32}_{-1.18}$}
\newcommand{\tessythrees}{1.25$^{+0.04}_{-0.04}$}
\newcommand{\tessythreeGPsigma}{4.19$^{+2.18}_{-1.13}$}
\newcommand{\tessythreeGPrho}{2.28$^{+1.25}_{-0.71}$}
\newcommand{\ASTEPonests}{5.49$^{+0.02}_{-0.01}$}
\newcommand{\ASTEPonestGPsigma}{0.59$^{+1.23}_{-0.46}$}
\newcommand{\ASTEPonestGPrho}{4.58$^{+181.50}_{-4.45}$}
\newcommand{\ASTEPtwonds}{5.49$^{+0.03}_{-0.01}$}
\newcommand{\ASTEPtwondGPsigma}{2.29$^{+4.47}_{-1.15}$}
\newcommand{\ASTEPtwondGPrho}{22.80$^{+281.40}_{-21.57}$}
\newcommand{\LCOBs}{5.50$^{+0.03}_{-0.02}$}
\newcommand{\LCOBGPsigma}{0.18$^{+0.56}_{-0.13}$}
\newcommand{\LCOBGPrho}{2.55$^{+155.36}_{-2.53}$}

\usepackage{xcolor}
\newcommand{\changes}[1]{\textcolor{black}{#1}}

\received{July 5, 2021}
\revised{\today}
\accepted{}
\submitjournal{AAS Journals}

\shorttitle{TOI-3362b: A Proto-Hot Jupiter Undergoing HEM}
\shortauthors{Dong, Huang, Zhou et al.}

\begin{document}

\title{TOI-3362b: A Proto-Hot Jupiter Undergoing High-Eccentricity Tidal Migration}

\newcommand{\PSUAA}{Department of Astronomy \& Astrophysics, 525 Davey Laboratory, The Pennsylvania State University, University Park, PA, 16802, USA}
\newcommand{\PSUCEHW}{Center for Exoplanets and Habitable Worlds, 525 Davey Laboratory, The Pennsylvania State University, University Park, PA, 16802, USA}
\newcommand{\USQAstro}{University of Southern Queensland, Centre for Astrophysics, West Street, Toowoomba, QLD 4350 Australia}
\newcommand{\CfA}{Center for Astrophysics \textbar \ Harvard \& Smithsonian, 60 Garden Street, Cambridge, MA 02138, USA}
\newcommand{\ArizonaAstro}{Department of Astronomy and Steward Observatory, University of Arizona, Tucson, AZ 85721, USA}
\newcommand{\IndianaAstro}{Department of Astronomy, Indiana University, Bloomington, IN 47405}
\newcommand{\MSUAstro}{Department of Physics and Astronomy, Michigan State University, East Lansing, MI 48824, USA}
\newcommand{\Birmingham}{School of Physics \& Astronomy, University of Birmingham, Edgbaston, Birmingham B15 2TT, United Kingdom}
\newcommand{\georgemason}{George Mason University, 4400 University Drive MS 3F3, Fairfax, VA, 22030 USA}
\newcommand{\oca}{Universit\'e C\^ote d'Azur, Observatoire de la C\^ote d'Azur, CNRS, Laboratoire Lagrange, Bd de l'Observatoire, CS 34229, 06304 Nice cedex 4, France}
\newcommand{\kavlimit}{Department of Physics and Kavli Institute for Astrophysics and Space Research, Massachusetts Institute of Technology, Cambridge, MA 02139, USA}
\newcommand{\PrincetonAstro}{Department of Astrophysical Sciences, Princeton University, 4 Ivy Lane, Princeton, NJ 08544, USA}

\correspondingauthor{Jiayin Dong}
\email{jdong@psu.edu}

\author[0000-0002-3610-6953]{Jiayin Dong}
\affiliation{\PSUAA}
\affiliation{\PSUCEHW}

\author[0000-0003-0918-7484]{Chelsea X. Huang}
\affiliation{\kavlimit}
\affiliation{\USQAstro}

\author[0000-0002-4891-3517]{George Zhou}
\affiliation{\USQAstro}

\author[0000-0001-9677-1296]{Rebekah I. Dawson} %consented
\affiliation{\PSUAA}
\affiliation{\PSUCEHW}

\author[0000-0001-8812-0565]{Joseph E. Rodriguez} %consented
\affiliation{\MSUAstro}

\author[0000-0003-3773-5142]{Jason D.\ Eastman} %consented
\affiliation{\CfA}

%karen.collins@cfa.harvard.edu
\author[0000-0001-6588-9574]{Karen A.\ Collins} %consented
\affiliation{\CfA}

%squinn@cfa.harvard.edu
\author[0000-0002-8964-8377]{Samuel N. Quinn} %consented
\affiliation{\CfA}

%shporer@mit.edu
\author[0000-0002-1836-3120]{Avi Shporer} %consented
\affiliation{\kavlimit}

%a.triaud@bham.ac.uk
\author[0000-0002-5510-8751]{Amaury H.M.J. Triaud} %consented
\affiliation{\Birmingham}

\author[0000-0002-7846-6981]{Songhu Wang} %consented
\affiliation{\IndianaAstro}

\author[0000-0002-9539-4203]{Thomas Beatty} %consented
\affiliation{\ArizonaAstro}

\author[0000-0002-0323-4828]{Jonathon Jackson} %consented
\affiliation{\PSUAA}
\affiliation{\PSUCEHW}

%kcolli3@gmu.edu; LCOGT   
\author[0000-0003-2781-3207]{Kevin I.\ Collins} %consented
\affiliation{\georgemason}

%Lyu.Abe@oca.eu; ASTEP team  
\author[0000-0002-0856-4527]{Lyu Abe} %consented
\affiliation{\oca}

%olga.suarez@oca.eu; ASTEP team
\author[0000-0002-3503-3617]{Olga Suarez} %consented
\affiliation{\oca}

% nicolas.crouzet@esa.int; ASTEP team
\author[0000-0001-7866-8738]{Nicolas Crouzet} %consented
\affiliation{European Space Agency (ESA), European Space Research and Technology Centre (ESTEC), Keplerlaan 1, 2201 AZ Noordwijk, The Netherlands}

% mekarnia@oca.eu; ASTEP team
\author[0000-0001-5000-7292]{Djamel M\'ekarnia} %consented
\affiliation{\oca}

% GXG831@student.bham.ac.uk; ASTEP team
\author[0000-0002-3937-630X]{Georgina Dransfield} %consented
\affiliation{\Birmingham}

%ejensen1@swarthmore.edu; SG1 member
\author[0000-0002-4625-7333]{Eric L.\ N.\ Jensen} %consented
\affiliation{Department of Physics \& Astronomy, Swarthmore College, Swarthmore PA 19081, USA}

%thestockdalefamily@bigpond.com; SG1 member
\author[0000-0003-2163-1437]{Chris Stockdale} %consented
\affiliation{Hazelwood Observatory, Australia}

%khalid.barkaoui@uliege.be
\author[0000-0003-1464-9276]{Khalid Barkaoui} %consented
\affiliation{Astrobiology Research Unit, Université de Liège, 19C Allée du 6 Août, 4000 Liège, Belgium}
\affiliation{Oukaimeden Observatory, High Energy Physics and Astrophysics Laboratory, Cadi Ayyad University, Marrakech, Morocco}

%alexis.heitzmann@usq.edu.au; MINERVA-Australis team
\author[0000-0002-8091-7526]{Alexis Heitzmann} %consented
\affil{\USQAstro}

%duncan.wright@usq.edu.au; MINERVA-Australis team
\author[0000-0001-7294-5386]{Duncan J. Wright} %consented
\affil{\USQAstro}

%Brett.Addison@usq.edu.au; MINERVA-Australis team
\author[0000-0003-3216-0626]{Brett C. Addison} %consented
\affil{\USQAstro}

%Rob.Wittenmyer@usq.edu.au; MINERVA-Australis team
\author[0000-0001-9957-9304]{Robert A. Wittenmyer} %consented
\affil{\USQAstro}

%Jack.Soutter@usq.edu.au; MINERVA-Australis team
\author[0000-0002-4876-8540]{Jack Okumura} %consented
\affil{\USQAstro}

%bpbowler@astro.as.utexas.edu; MINERVA-Australis team
\author[0000-0003-2649-2288]{Brendan P. Bowler} %consented
\affil{Department of Astronomy, The University of Texas at Austin, TX 78712, USA}

%jonti.horner@usq.edu.au; MINERVA-Australis team
\author[0000-0002-1160-7970]{Jonathan Horner} %consented
\affil{\USQAstro}

%skane@ucr.edu; MINERVA-Australis team
\author[0000-0002-7084-0529]{Stephen R. Kane} %consented
\affil{Department of Earth and Planetary Sciences, University of California, Riverside, CA 92521, USA}

%kielkopf@louisville.edu; MINERVA-Australis team
\author[0000-0003-0497-2651]{John Kielkopf} %consented
\affil{Department of Physics and Astronomy, University of Louisville, Louisville, KY 40292, USA}

%huigen@nju.edu.cn; MINERVA-Australis team
\author{Huigen Liu} %consented
\affil{School of Astronomy and Space Science, Key Laboratory of Modern Astronomy and Astrophysics in Ministry of Education, Nanjing University, Nanjing 210046, Jiangsu, China}

%pplavcha@gmu.edu; MINERVA-Australis team
\author[0000-0002-8864-1667]{Peter Plavchan} %consented
\affil{\georgemason}

%Matthew.Mengel@usq.edu.au; MINERVA-Australis team
\author[0000-0002-7830-6822]{Matthew W. Mengel} %consented
\affil{\USQAstro}

% TESS Architect 
\author[0000-0003-2058-6662]{George R. Ricker} %consented
\affil{\kavlimit}

\author[0000-0001-6763-6562]{Roland Vanderspek} %consented
\affiliation{\kavlimit}

\author[0000-0001-9911-7388]{David W. Latham} %consented
\affil{\CfA}

\author[0000-0002-6892-6948]{S. Seager} %consented
\affil{Department of Earth, Atmospheric, and Planetary Sciences, Massachusetts Institute of Technology, Cambridge, MA 02139, USA}
\affil{\kavlimit}
\affil{Department of Aeronautics and Astronautics, Massachusetts Institute of Technology, Cambridge, MA 02139, USA}

\author[0000-0002-4265-047X]{Joshua N.\ Winn} %consented
\affiliation{\PrincetonAstro}

\author[0000-0002-4715-9460]{Jon M. Jenkins} %consented
\affiliation{NASA Ames Research Center, Moffett Field, CA 94035, USA}

\author[0000-0002-8035-4778]{Jessie L. Christiansen} %consented
\affil{Caltech/IPAC-NASA Exoplanet Science Institute, Pasadena, CA 91125}

\author[0000-0001-8120-7457]{Martin Paegert} %consented
\affiliation{\CfA}

\begin{abstract}
High-eccentricity tidal migration is a possible way for giant planets to be emplaced in short-period orbits. If it commonly operates, one would expect to catch proto-Hot Jupiters on highly elliptical orbits that are undergoing high-eccentricity tidal migration. As of yet, few such systems have been discovered. Here, we introduce \toib (\ticb), an 18.1-day, 5 \mj{} planet orbiting a main-sequence F-type star that is likely undergoing high-eccentricity tidal migration. The orbital eccentricity is \ecc. With a semi-major axis of \aplt{} au, the planet's orbit is expected to shrink to a final orbital radius of \afinal{} au after complete tidal circularization. Several mechanisms could explain the extreme value of the planet's eccentricity, such as planet-planet scattering and secular interactions. Such hypotheses can be tested with follow-up observations of the system, e.g., measuring the stellar obliquity and \changes{searching for companions in the system with precise, long-term radial velocity observations}. The variation in the planet's equilibrium temperature as it orbits the host star and the tidal heating at periapse make this planet an intriguing target for atmospheric modeling and observation. Because the planet's orbital period of 18.1 days is near the limit of \TESS's period sensitivity, even a few such discoveries suggest that proto-Hot Jupiters may be quite common.
\end{abstract}

\keywords{Exoplanets (498), Hot Jupiters (753), Transit photometry (1709), Radial velocity (1332)}

\section{Introduction} \label{sec:intro}
High-eccentricity tidal migration has long been proposed to explain the existence of close-in giant planets. In this scenario, a giant planet is formed several au away from its host star. \changes{Its orbital eccentricity is excited to a high value, perhaps via planet-planet scattering \citep[e.g.,][]{rasi96, chat08, naga08} or secular interactions, such as von Zeipel-Lidov-Kozai oscillations (e.g., \citealt{vonzeipel1910, lido62, koza62, wu03}; see \citealt{naoz16} for a review)} and secular chaos \citep[e.g.,][]{wu11}. 
The orbit loses energy due to tidal dissipation as the planet approaches the star near periapse. Once the planet decouples from its perturber, the planet's orbital angular momentum is expected to be conserved (i.e, $L = M_p\sqrt{G M_\star a_p (1-e_p^2)}$ is constant). As a consequence, the initially Cold \changes{planet} migrates inwards while its orbit circularizes, following the track $a_p(1-e_p^2) = \textrm{constant} = a_{\rm final}$. The final orbital radius of planet $a_{\rm final}$ depends on the planet's initial semi-major axis and eccentricity. The tidal circularization timescale $\tau_{\rm c}$ has a strong dependence on $a_{\rm final}$ (\changes{e.g.}, $\tau_{\rm c} \propto a_{\rm final}^8$; \citealt{eggl89, eggl01, hans10}) and therefore, only Cold \changes{planets} with sufficiently small $a_{\rm final}$ could possibly circularize during the system's lifetime.

If close-in giant planets are products of high-eccentricity tidal migration, we would expect to catch at least some of them on the tidal circularization track (e.g., \citealt{socr12}). For example, a planet tidally migrating to $a_{\rm final} = 0.05$ au will have an eccentricity of 0.86 when it is at 0.2 au. \changes{As of yet, few close-in giant planets have been observed with such extreme eccentricities and sufficiently small $a_{\rm final}$ (i.e., $a_{\rm final} \lessapprox 0.05$ au; \citealt{daws15}). HD 80606b ($a_{\rm p} = 0.46$ au, $e_{\rm p} = 0.93$; \citealt{naef01}) might be the most impressive example of a proto-Hot Jupiter undergoing high-eccentricity tidal migration to be a Hot Jupiter. The planet's distant stellar companion suggests stellar Kozai cycles as the possible mechanism to excite its extreme eccentricity \citep{wu03}. HAT-P-2b (HD 147506b; $a_{\rm p} = 0.068$ au, $e_{\rm p} = 0.52$; \citealt{bako07}), which has a less extreme eccentricity, could be another example of a proto-Hot Jupiter undergoing tidal migration and circularization. The long-term radial-velocity observations of the system reveals the presence of a substellar companion \citep{lewi13}, also suggesting the stellar Kozai mechanism as the source of the eccentricity excitation. For planets with lower masses, Kepler-1656b ($a_{\rm p} = 0.20$ au, $e_{\rm p} = 0.84$, $M_{\rm p} = 48 M_\Earth$; \citealt{brad18}) is an extraordinary example. The high physical density of the planet, along with the extreme eccentricity, might suggest planet-planet interactions such as collision and scattering in the planet's dynamical history.}

In this work, we introduce \toib (\ticb, 2MASS J10235624-5650353), a transiting proto-Hot Jupiter that is also likely undergoing high-eccentricity tidal circularization. \toib was first identified by a systematic search for Warm Jupiters in the Southern Ecliptic Hemisphere in the Year 1 of the \TESS Full-Frame Images data \citep{dong21}. The unusually short duration of the transits compared to the expected duration for a circular orbit \citep[e.g.,][]{daws12} pointed to its high eccentricity. Ground-based follow-up observations presented here validate the planet, break the degeneracy between the eccentricity and argument of periapse, and constrain the planet's mass.

In Section~\ref{sec:observation}, we describe the \TESS and the ground-based photometric and spectroscopic follow-up observations of the target by \astep, \lco, \chiron, and \minerva. In Section~\ref{sec:results}, we fit a model to determine the stellar and planetary parameters. In Section~\ref{sec:discussion}, we discuss the dynamical implications and motivate further investigations of the system.

\section{Observations} \label{sec:observation}
Here we describe the \TESS photometry in Section~\ref{subsec:tess}, ground-based transit photometry by \astep and \lco \citep{Brown:2013} in Section~\ref{subsec:gb_transit}, and ground-based spectroscopic observations by \chiron and \minerva in Section~\ref{subsec:gb_rv}. 

\subsection{TESS Photometry} \label{subsec:tess}

\begin{figure*}
    \centering
    \includegraphics{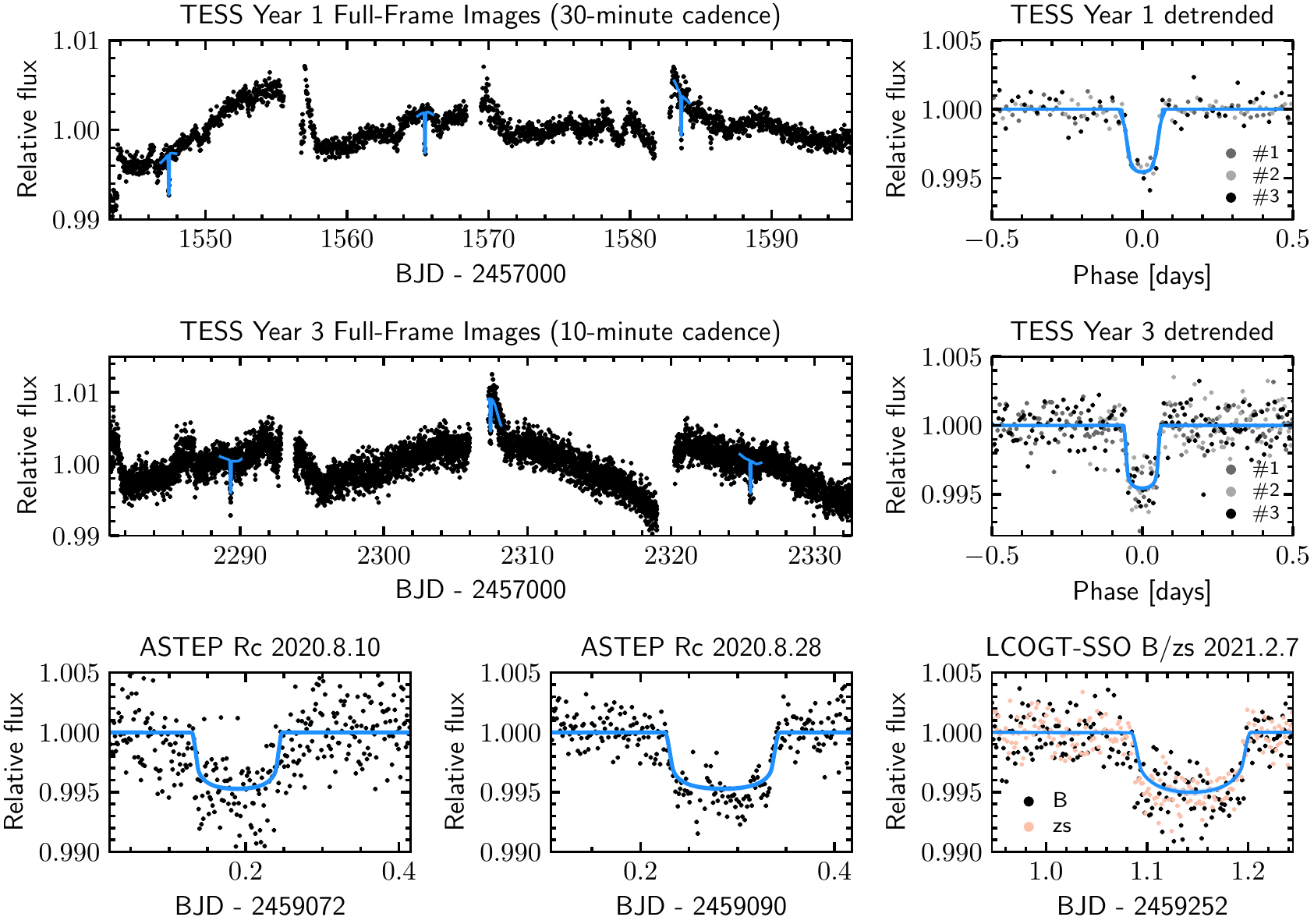}
    \caption{\TESS and ground-based transit photometry of \toib overplotted with best modeled light curves. Both raw (upper and middle left panels) and folded-detrended (upper and middle right panels) \TESS Year 1 (30-minute cadence) and Year 3 (10-minute cadence) light curves are presented. 
    Three detrended, ground-based light curves observed with \astep and \lco are presented. The \lco transit was detected in two different filters (B and \filterzs{}) with no obvious transit depth variation detected.}
    \label{fig:lcs}
\end{figure*}

\toi (\tic) was observed by \TESS with 30-minute cadence during Sectors 9 and 10 of its primary mission (2019-Feb-28 to 2019-Apr-22), and with 10-minute cadence during Sector 36 and 37 of its first extended mission (2021-Mar-07 to 2021-Apr-28). During a systematic search for warm, large planets using Quick Look Pipeline light curves derived from \TESS primary Full Frame Images \citep[FFIs;][]{huan20b, huan20c, dong21}, we detected a candidate transit signal at period of 18.134 days, with signal to pink noise (i.e., $1/f$ frequency noise) ratio of 22.7. 
Three transits of \toib were observed during the \TESS primary mission. Preliminary analysis showed that the stellar density ratio $\rho_{\rm circ}/\rho_\star$, where $\rho_{\rm circ}$ is the inferred stellar density from the light curves assuming a circular orbit and $\rho_\star$ is from isochrone fitting, was about 18 (i.e., much larger than 1), and therefore the planetary candidate is highly likely to be on an eccentric orbit \citep{dong14}. Here we use early released TESS Image CAlibrator (TICA) High Level Science Product (HLSP) FFIs \citep{faus20} in the extended mission to derive the 10-minute cadence light curves. The method we employ is similar to that used to derive the standard QLP light curves. Three additional transits are observed. 
We show the raw and detrended light curves in Figure \ref{fig:lcs}. The light curves are detrended using a Matern-3/2 Gaussian Process (GP) kernel \citep{exoplanet:foremanmackey17, exoplanet:foremanmackey18}. 
There are no \TESS spacecraft events impacting any of the transits. The second transit in the Year 3 data has relatively high level of noise since it occurred at the beginning of a \TESS orbit. We perform the light-curve fit with and without the second transit in Year 3 and find similar planet-star radius ratio posteriors. See Section~\ref{subsec:lc-only} for more details.

\subsection{Ground-based Transit Photometry} \label{subsec:gb_transit}

We used the $\mathtt{TESS Transit Finder}$, which is a customized version of the $\mathtt{Tapir}$ software package \citep{Jensen:2013}, to schedule our transit observations. We observed three transits as part of the \TESS Follow-up Observing Program. The detrended light curves from these observations can all be found on the ExoFOP-TESS website.\footnote{\href{https://exofop.ipac.caltech.edu/tess/}{https://exofop.ipac.caltech.edu/tess/}}

The Antarctica Search for Transiting ExoPlanets (ASTEP) program observed two transits on the nights of UT 2020 August 10 and UT 2020 August 28 \citep{guil15, meka16}. The \SI{0.4}{\m} telescope is equipped with an FLI Proline science camera with a KAF-16801E, 4096 × 4096 front-illuminated CCD and is located on the East Antarctic plateau. The non-filtered red science channel is similar to the \filterrc{} band in transmission. The first transit was observed in a \ang{;;11.2} aperture including a neighboring star 3.3 magnitude fainter than \toi. The second transit was observed in a \ang{;;10.1} aperture. The estimated transit depths from both transits are consistent with the \TESS observations.  

A full transit of \toib was observed in the Pan-STARSS \filterzs{} and Bessell \emph{B} band on \ut{} 2021 February 7 using a \SI{1.0}{\m} telescope at the Las Cumbres Observatory Global Telescope (LCOGT; \citealt{Brown:2013}) Siding Spring Observatory (SSO) node in New South Wales, Australia. The LCOGT observations were calibrated with the standard BANZAI pipeline, and the light curves were extracted using $\mathtt{AstroImageJ}$ \citep[AIJ;][]{Collins:2017}. 
The observation used an uncontaminated \ang{;;3.5} aperture for the \filterzs{} and a \ang{;;4.67} aperture for the \emph{B} band and recovered the expected transit signal. The transit depths from the two bands show no strong chromaticity.

\subsection{Spectroscopic Observations} \label{subsec:gb_rv}

Spectroscopic follow-up observations of \toi were conducted to measure the orbit and mass of the transiting planet, and to refine the atmospheric parameters of the host star. We obtained 21 spectra using the \chiron echelle spectrograph \citep{toko13} on the SMARTS \SI{1.5}{\m} telescope located at the Cerro Tololo Inter-American Observatory (CTIO), Chile, between \ut{} 2021 April 15 and 2021 May 12. \chiron is a high resolution echelle spectrograph fed via an image slicer and a fiber bundle, with a resolving power of $R \approx \num{80000}$ over the wavelength range of \SIrange[range-phrase=--]{4100}{8700}{\angstrom}. The wavelength solution is provided via Th-Ar hollow-cathode lamp exposures that were taken after each science exposure. The spectra were extracted via the official CHIRON pipeline \citep{toko13}.
The radial velocities were derived as per the procedure described in \citet{zhou20}. Briefly, we derived a line broadening profile for each spectrum using a least-squares deconvolution \citep{donati97} between the observed spectra and a synthetic nonrotating spectral templates generated via the ATLAS9 stellar models \citep{cast14}. The line broadening profiles were fitted via a convolution of the rotational, macroturbulent, and instrumental broadening kernels, yielding the radial velocity shift and the rotational broadening velocity. To derive spectroscopic atmospheric parameters for \toi, we match the CHIRON spectra against an observed spectral library that has been classified by the Spectral Classification Pipeline \citep{buch12}, following the procedure described in \citet{zhou20}. We measure a spectroscopic effective temperature of $T_\mathrm{eff} = 6532^{+88}_{-86}$\,K, surface gravity $\log g = 4.072^{+0.032}_{-0.034}$ dex, and bulk metallicity $\mathrm{[M/H]} = 0.017^{+0.057}_{-0.049}$ dex. The star has a stellar type of F5V. 

We also obtained nine observations of \toi between 2021 May 16 and 2021 May 30 using the Minerva-Australis telescope array \citep{addison19}, located at Mt.\ Kent Observatory, Australia. Minerva-Australis is an array of four identical \SI{0.7}{\m} telescopes linked via fiber feeds to a single KiwiSpec echelle spectrograph, at a spectral resolving power of $R \approx \num{80000}$ over the wavelength region of \SIrange[range-phrase=--]{5000}{6300}{\angstrom}. The array is wholly dedicated to radial-velocity follow-up of \TESS planet candidates \citep[e.g.][]{niel19, brah20, addi21}.
Simultaneous wavelength calibration is provided via two calibration fibers illuminated by a quartz lamp through an iodine cell. The spectra were extracted for each telescope individually, and the radial velocities extracted via the same techniques as those described above for the CHIRON observations. We note, though, that due to throughput issues, only velocities from Telescope 1 of the MINERVA-Australis array were adopted for this analysis; the remaining telescopes yielding velocity uncertainties too large to contribute meaningfully to the orbit detection.

\section{Results} \label{sec:results}
Here we describe models to infer the stellar and planetary parameters. In Section~\ref{subsec:stellar}, we present the stellar isochrone and Spectral Energy Distribution (SED) fitting results. In Section~\ref{subsec:lc-only}, we show the transit-only model and report the transit-timing variation analysis and planet's eccentricity inferred purely from light curves. In Section~\ref{subsec:joint}, we describe the transit and radial-velocity joint model and present the planetary parameters.

\subsection{Stellar Properties} \label{subsec:stellar}

We used the exoplanet fitting suite, $\mathtt{EXOFASTv2}$ \citep{Eastman:2013, Eastman:2019}, to perform a fit of the Spectral Energy Distribution (SED) for \toi. We combined the available Gaia ($G$, $B_P$, and $R_P$, \citealp{Gaia:2018}), 2MASS ($J_{2\rm{M}}, H_{2\rm{M}}$, and $K_{S, 2\rm{M}}$ \citealp{Cutri:2003}), and WISE (W1 and W2, \citealp{Zacharias:2017}) photometry with Gaussian priors on the metallicity (0.002$\pm$0.070) and T$_{\rm eff}$ (6513$\pm$100K) from spectroscopy (see \S\ref{subsec:gb_rv}), and Gaia DR2 parallax (2.7265$\pm$0.03952 mas, corrected for the -30 $\mu$as offset reported by \citet{Lindegren:2018}). We also placed an upper limit on the line of sight extinction of 4.1844 mag from \citet{Schlegel:1998} \& \citet{Schlafly:2011}. Within the fit, the MESA Isochrones and Stellar Tracks (MIST) stellar evolution models \citep{Choi:2016, Dotter:2016} are used to provide better estimates for the host star parameters. We also put a lower bound on the age of the host star of 100 Myr since we do not see any signs of youth in the photometry or spectroscopy (e.g., from the stellar rotation period and lithium abundance). \tic\ has a mass of $1.445^{+0.069}_{-0.073}$ M$_\odot$, a radius of $1.830^{+0.055}_{-0.053}$ R$_\odot$, and an age of $2.14^{+0.66}_{-0.52}$ Gyr. See Table \ref{tab:parameters} for a complete list of the results from this analysis. 

We also calculated the stellar density and radius from isochrone fitting with the Dartmouth \citep{dott08} stellar evolution models. The approach -- as described by \citet{daws15} -- fits the stellar effective temperature, metallicity, \emph{Gaia} DR2 parallax, and \emph{Gaia} apparent $g$ magnitude \citep{gaia16,gaia18}, where the stellar temperature, metallicity, and uncertainties are inferred from the CHIRON spectra. We applied the systematic correction to \emph{Gaia} parallaxes from \citet{stas18b}. The values derived from the Dartmouth model are consistent with the MIST model and have similar uncertainties.

\subsection{Transit-only Model} \label{subsec:lc-only}
The \TESS and ground-based transit photometry have been described in Section~\ref{subsec:tess} and \ref{subsec:gb_transit} and shown in Figure~\ref{fig:lcs}. We only model \TESS light curves around each transit to roughly 6 times the transit duration to reduce computational expense. We use a quadratic limb darkening transit model \citep{mand02, exoplanet:kipping13} plus a Matern-3/2 GP kernel with a white-noise term to model the light curves \citep{exoplanet:foremanmackey17, exoplanet:foremanmackey18}. We use informative limb darkening coefficients from \cite{clar17} for transits in different filters ($\TESS$, \filterrc{}, \emph{B}, \filterzs{}).
\TESS Year 1 observations, Year 3 observations, and each of the three ground-based observations have their own GP kernels to account for differences in variability captured by different wavelengths, instruments, and cadences. The stellar density $\rho_{\textrm{circ}}$ is modeled assuming the planet has a circular orbit. We later compare the marginalized $\rho_{\textrm{circ}}$ to $\rho_{\star}$, the stellar density derived from the isochrone fitting, to constrain the planet's eccentricity. To characterize the transit-timing variation signal of the planet, we model the mid-transit times $T_{1..N}$ individually. 
All free parameters in the transit-only model are 
\begin{align}
    \big\{&\rho_{\textrm{circ}}, b, r_{\textrm{p}}/r_{\star}, T_{1..N},\nonumber\\
    &u_{0..1, \TESS}, u_{0..1, Rc}, u_{0..1, B}, u_{0..1, \filterzs},\nonumber\\
    &s_{\rm TESS, Y1}, \rho_{\rm TESS, Y1}, \sigma_{\rm TESS, Y1}\nonumber\\
    &s_{\rm TESS, Y3}, \rho_{\rm TESS, Y3}, \sigma_{\rm TESS, Y3}\nonumber\\
    &s_{\rm ASTEP, 1}, \rho_{\rm ASTEP, 1}, \sigma_{\rm ASTEP, 1}\nonumber\\
    &s_{\rm ASTEP, 2}, \rho_{\rm ASTEP, 2}, \sigma_{\rm ASTEP, 2}\nonumber\\
    &s_{\rm LCOGT}, \rho_{\rm LCOGT}, \sigma_{\rm LCOGT}\big\},
\end{align}
where $\rho_{\textrm{circ}}$ is the stellar density assuming a circular orbit, $b$ is the impact parameter, $r_{\textrm{p}}/r_{\star}$ is the planet-star radius ratio, $u_{0..1}$ are the quadratic limb darkening coefficients, and $s$ is the photometric white noise. The Matern-3/2 GP kernel follows $\mathcal{K}(\tau) = \sigma^2 (1 + \sqrt{3}\tau/\rho)\exp{(-\sqrt{3}\tau/\rho)}$, where $\sigma$ presents the amplitude of variability and $\rho$ presents the timescale. The priors used in this model are the same as the ones listed in Table 2, \cite{dong21}.

\begin{figure}
    \centering
    \includegraphics{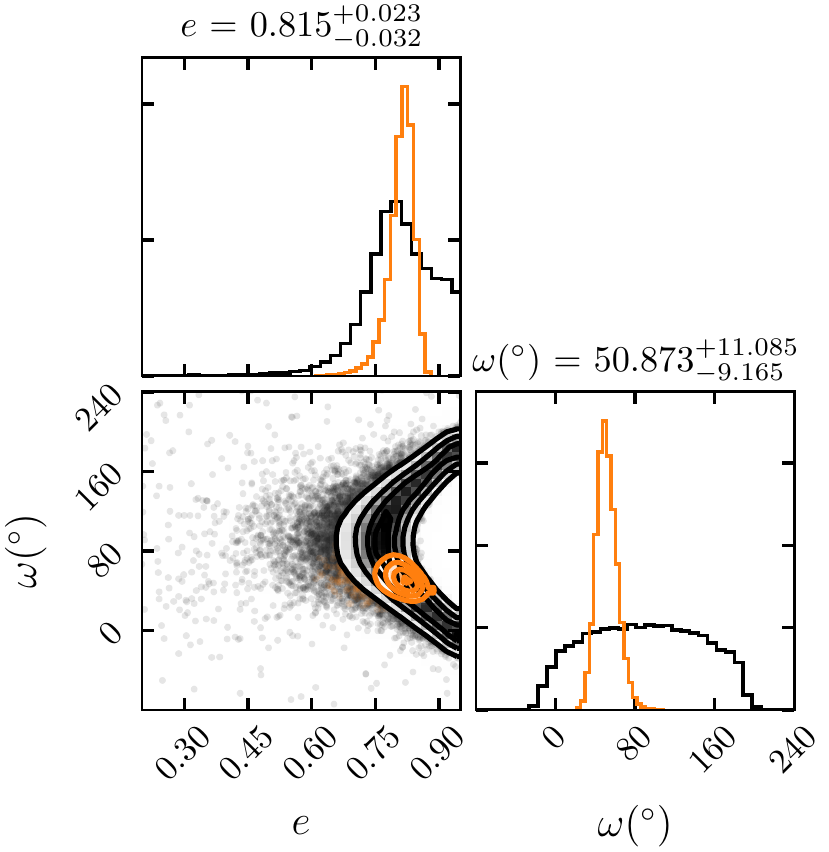}
    \caption{Joint posterior distributions of the eccentricity and argument of periapse for \toib. Values inferred purely from the ``photoeccentric" effect are colored in black and from the transit$+$RV joint fit in orange. \changes{The contours are 0.5, 1, 1.5, and 2 sigma levels. Radial-velocity measurements confirm the extreme eccentricity of \toib and provide much tighter constraints.}}
    \label{fig:ecc}
\end{figure}

We use the $\mathtt{exoplanet}$ package \citep{fore21} for the light curve fits. Four chains are sampled using the Markov Chain Monte Carlo (MCMC) technique with gradient-based proposals \citep{neal12, hoff11, beta17}. Each chain contains 50,000 tuning steps and 3000 sampling steps with a target accept rate of 0.95. We assess MCMC convergence using the summary statistics, e.g., Gelman-Rubin diagnostic ($\hat{\mathcal{R}} < 1.1$ for convergence), trace plots, and corner plots \citep{corner} of the marginal joint distributions. The $\hat{\mathcal{R}}$ values for our model parameters are all less than 1.001 by the end of the sampling. All four chains are combined to get the posteriors.

We use the medians and 1$\sigma$ uncertainties of the mid-transit time posteriors to perform a least-square fit to a linear line. The orbital period and the conjunction time of the transiting planet are inferred. The $O$-$C$ residuals, which are calculated by subtracting linear ephemerides from observed mid-transit times, present no significant transit-timing variations. We can rule out any level of transit-timing variability greater than $\sim$5 minutes.

From the stellar density posteriors $\rho_{\textrm{circ}}$ and $\rho_{\star}$, we infer the eccentricity $e$ and argument of periapse $\omega$ of the planet, shown as the black histograms and contours in Figure~\ref{fig:ecc}. The stellar densities suggest the planet is on a highly elliptical orbit.

\subsection{Transit-only model with EXOFASTv2}
\label{subsec:exofasttran}
\changes{For a consistency check, we also modeled the \TESS light curve and ground based transits, along with the stellar model described in \S \ref{subsec:stellar} and including planetary eccentricity directly, using $\mathtt{ EXOFASTv2}$ \citep{Eastman:2019}.}

\changes{We used $\mathtt{keplerspline}$ from \citet{Vanderburg:2014} to account for the long term variability of the \TESS lightcurves, and averaged 10 model data points over the 30-minute Sectors 9 and 10 exposures, and 4 model data points over the 10-minute Sectors 36 and 37 exposures. We also fit a dilution term to the \TESS lightcurves to account for poor background subtraction \citep{Burt:2020}. In practice, fitting the dilution term means the transit depth is defined by the ground based lightcurves.}

\changes{The stellar model therefore constrains the stellar density, and the transit duration then constrains the eccentricity.}

\changes{Most of our values were consistent with the results shown in \S \ref{subsec:lc-only}. In particular, our eccentricity was within $0.05 \sigma$. The worst agreement was in our fractional depth which we measured as $R_P/R_* = 0.0685^{+0.0012}_{-0.0010}$ -- 2.3$\sigma$ discrepant, likely because of the different detrending methods.}

\begin{table}
\scriptsize
\setlength{\tabcolsep}{1pt}
\renewcommand{\arraystretch}{0.85}
\centering
\caption{Median values and 68\% confidence intervals for the stellar and planetary parameters of \toib. \label{tab:parameters}}
\begin{tabular}{llccc}
  \hline
  \hline
Parameter & Units & Values \\
\hline\\\multicolumn{2}{l}{Stellar Parameters:$^{1}$}&\smallskip\\
~~~~$M_*$\dotfill &Mass (\msun)\dotfill &$1.445^{+0.069}_{-0.073}$\\
~~~~$R_*$\dotfill &Radius (\rsun)\dotfill &$1.830^{+0.055}_{-0.053}$\\
~~~~$L_*$\dotfill &Luminosity (\lsun)\dotfill &$5.48^{+0.37}_{-0.32}$\\
~~~~$\rho_*$\dotfill &Density (cgs)\dotfill &$0.331^{+0.035}_{-0.032}$\\
~~~~$\log{g}$\dotfill &Surface gravity (cgs)\dotfill &$4.072^{+0.032}_{-0.034}$\\
~~~~$T_{\rm eff}$\dotfill &Effective temperature (K)\dotfill &$6532^{+88}_{-86}$\\
~~~~$[{\rm M/H}]$\dotfill & Bulk metallicity (dex)\dotfill &$0.017^{+0.057}_{-0.049}$\\
~~~~$v\sin I_\star $\dotfill & Rotational line broadening ($\mathrm{km\,s}^{-1}$)\dotfill &$20.0\pm4.1$\\
~~~~$Age$\dotfill &Age (Gyr)\dotfill &$2.14^{+0.66}_{-0.52}$\\
~~~~$EEP$\dotfill &Equal evolutionary phase$^{2}$ \dotfill &$369^{+22}_{-19}$\\
~~~~$A_V$\dotfill &V-band extinction (mag)\dotfill &$0.098^{+0.062}_{-0.056}$\\
~~~~$\sigma_{\mathrm{SED}}$\dotfill &SED photometry error scaling \dotfill &$0.50^{+0.18}_{-0.11}$\\
~~~~$\varpi$\dotfill &Parallax (mas)\dotfill &$2.724^{+0.040}_{-0.039}$\\
~~~~$G$\dotfill &Gaia $G$ magnitude\dotfill &$10.705\pm0.020$\\
~~~~$B_{\mathrm{P}}$\dotfill &Gaia $B_{\mathrm{P}}$ magnitude\dotfill &$10.948\pm0.020$\\
~~~~$R_{\mathrm{P}}$\dotfill &Gaia $R_{\mathrm{P}}$ magnitude\dotfill &$10.335\pm0.020$\\
~~~~$J$\dotfill &2MASS $J$ magnitude \dotfill &$9.944\pm0.024$\\
~~~~$H$\dotfill &2MASS $H$ magnitude \dotfill &$9.719\pm0.022$\\
~~~~$K_{\mathrm{S}}$\dotfill &2MASS $K_{\mathrm{S}}$ magnitude \dotfill &$9.693\pm0.023$\\
~~~~WISE1\dotfill & WISE1 magnitude\dotfill &$9.651\pm0.030$\\
~~~~WISE2\dotfill & WISE2 magnitude\dotfill &$9.678\pm0.030$\\
~~~~WISE3\dotfill & WISE3 magnitude\dotfill &$9.643\pm0.063$\\
~~~~WISE4\dotfill & WISE4 magnitude\dotfill &$9.427\pm0.516$\\
\\
\hline\\\multicolumn{2}{l}{Planetary Parameters (joint model):}&\smallskip\\
~~~~$P$\dotfill &Period (days)\dotfill & \period\\
~~~~$T_C$\dotfill &Mid-transit time (BJD-2457000)\dotfill & \midt\\
~~~~$b$\dotfill &Transit impact parameter \dotfill & \bp\\
~~~~$\delta$\dotfill &Transit depth \dotfill & \depth\\
~~~~$M_P$\dotfill &Mass (\mj)\dotfill & \mpl\\
~~~~$R_P$\dotfill &Radius (\rj)\dotfill & \rpl\\
~~~~$a$\dotfill &Semi-major axis (au)\dotfill & \aplt\\
~~~~$i$\dotfill &Inclination ($^\circ$)\dotfill & \inc\\
~~~~$e$\dotfill &Eccentricity \dotfill & \ecc\\
~~~~$\omega$\dotfill &Argument of periapse ($^\circ$)\dotfill & \argperi\\
\\
\hline\\\multicolumn{2}{l}{Instrumental Parameters:} & $u_0$ & $u_1$\smallskip\\
~~~~$u_{\TESS}$\dotfill &Limb-darkening coefficients\dotfill & \utessa & \utessb\\
~~~~$u_{Rc}$\dotfill &Limb-darkening coefficients\dotfill & \uRca & \uRcb\\
~~~~$u_{B}$\dotfill &Limb-darkening coefficients\dotfill & \uBa & \uBb\\
~~~~$u_{\filterzs}$\dotfill &Limb-darkening coefficients\dotfill & \uzsa & \uzsb\\
\smallskip\\
& & $s$ [ppt] & $\rho$ [days] & $\sigma$ [ppt]\smallskip\\
~~~~\TESS, Y1\dotfill &Gaussian Process parameters\dotfill & \tessyones & \tessyoneGPrho & \tessyoneGPsigma\\
~~~~\TESS, Y3\dotfill &Gaussian Process parameters\dotfill & \tessythrees & \tessythreeGPrho & \tessythreeGPsigma\\
~~~~\astep, 1\dotfill &Gaussian Process parameters\dotfill & \ASTEPonests & \ASTEPonestGPrho & \ASTEPonestGPsigma\\
~~~~\astep, 2\dotfill &Gaussian Process parameters\dotfill & \ASTEPtwonds & \ASTEPtwondGPrho & \ASTEPtwondGPsigma\\
~~~~LCOGT \dotfill &Gaussian Process parameters\dotfill & \LCOBs & \LCOBGPrho & \LCOBGPsigma\\
\smallskip\\
& & \chiron & Minerva\\
~~~~$\gamma_{\rm rel}$\dotfill &Relative RV offset ($\mathrm{m\,s}^{-1}$)\dotfill & \chironoffset & \minervaoffset\\
~~~~$\sigma_J$\dotfill &RV jitter ($\mathrm{m\,s}^{-1}$)\dotfill & \chironjitter & \minervajitter\\
~~~~$k_{\rm bkg}$\dotfill & Radial velocity baseline ($\mathrm{m\,s}^{-1}{\rm day}^{-1}$)\dotfill & \chirontrend & \minervatrend\\
\smallskip\\
\hline
\end{tabular}
\begin{flushleft}
 \footnotesize{ \textbf{\textsc{NOTES:}\\}
$^1$The stellar parameters are derived purely from the MIST and SED fitting. See Table 3 in \citet{Eastman:2019} for a detailed description of stellar parameters.\\ 
$^2$Corresponds to static points in a star's evolutionary history. See \S2 in \citet{Dotter:2016}.\\
}
\end{flushleft}
\end{table}

\subsection{Joint Transit and Radial Velocity Modeling} \label{subsec:joint}

Here we combine the photometry with ground-based spectroscopic observations, as described in Section~\ref{subsec:gb_rv}, to infer the planet's mass and put tighter constraint on the eccentricity. Since the planet shows no evidence of transit-timing variations, we directly model its orbital period and conjunction time. The priors of these values \changes{follow normal distributions with the medians of 18.1 days and 2458529.3 BJD, respectively and uncertainties of 0.1 days}.

The joint model has the parameters mentioned in the transit-only model but also includes planet mass $M_p$, orbital eccentricity $e$, and argument of periapse $\omega$. Only the stellar density was modeled in the transit-only analysis. To jointly fit the radial velocity data, we will also need the stellar mass. Picking two parameters out of $\rho_\star$, $M_\star$, and $R_\star$ will be sufficient. We choose $M_\star$ and $R_\star$  for the joint model since the pair show little covariance in the stellar-fit posteriors discussed in Section~\ref{subsec:stellar}. The stellar-fit posteriors are used as the priors for $M_\star$ and $R_\star$ in the joint model. To account for the instrumentation offsets and systematics, both \chiron and \minerva are given an additive offset parameter and a log-Normal radial velocity jitter term. We also include a linear function of time for both instruments as a baseline model for stellar activity.

\begin{figure*}
    \centering
    \includegraphics{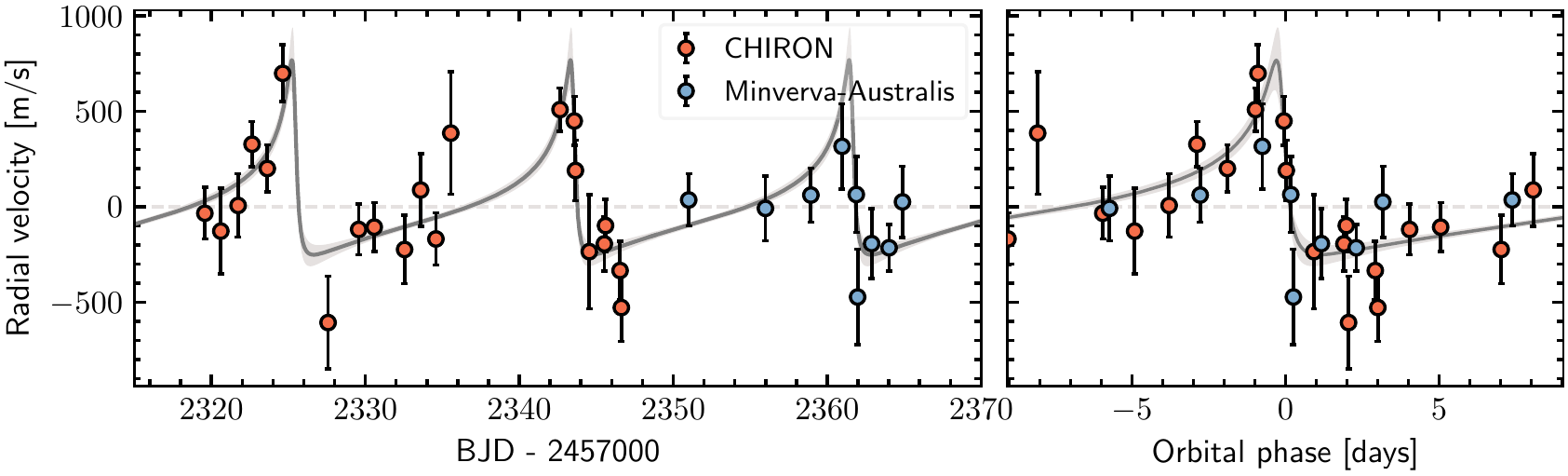}
    \caption{Radial velocity observations of \toi by \chiron colored in orange and \minerva colored in blue. The instrumental offsets and linear background trends have been abstracted from \chiron and \minerva data. The phased radial velocity is plotted in the right panel, where the planet transits at $t = 0$. The median and 1$\sigma$ uncertainty of the planetary radial velocity signal are shown in grey.}
    \label{fig:rvs}
\end{figure*}

We use the $\mathtt{exoplanet}$ package for the joint fit, following the same sampling and post-analysis procedures as described in Section~\ref{subsec:lc-only}. No significant background trends are found in \chiron and \minerva observations. A summary of the posteriors can be found in Table~\ref{tab:parameters}. Best modeled light curves and radial velocity orbit can be found in Figure~\ref{fig:lcs} and \ref{fig:rvs}, respectively. 

\begin{figure*}
    \centering
    \hspace*{-0.1cm}
    \includegraphics{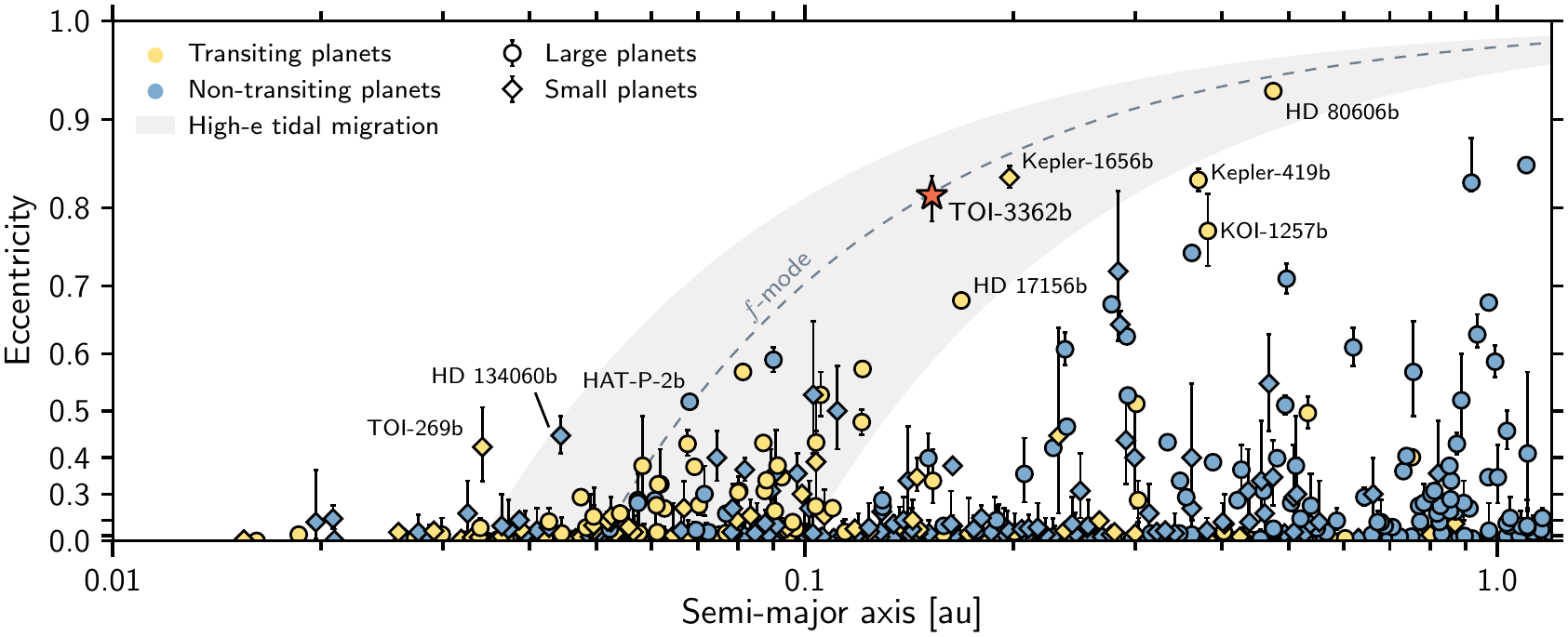}
    \caption{\changes{Eccentricity versus semi-major axis for all confirmed planets less than 13 Jupiter-mass with an orbital period less than 365 days. Data are extracted from the NASA Exoplanet Archive (DOI 10.26133/NEA12) as of July 26th, 2021.} The $y$-axis is scaled to $e^2$ to emphasize non-circular planets. The $x$-axis is in log scale. The grey region with a final semi-major axis range of 0.034--0.1 au is set by the Roche limit and the tidal circularization timescale, respectively. \changes{Planets above the dashed grey line could have their $f$-mode tidal dissipation excited to speed up the migration. Transiting planets are labeled in yellow and non-transiting planets are labeled in blue. Large planets ($R_p > 6 R_\Earth$ and/or $M_p > 100 M_\Earth$) are labeled as circles and small planets  ($R_p < 6 R_\Earth$ and/or $M_p < 100 M_\Earth$) are labeled as diamonds. To help with the interpretation, we only plot planets with well-constrained eccentricities, which are planets with eccentricity uncertainties less than 50\% of the measured eccentricities or uncertainties less than 0.2 for those on nearly circular orbit ($e <$ 0.2).}}
    \label{fig:hem}
\end{figure*}

\section{Discussion and Future Work} \label{sec:discussion}

The inferred semi-major axis and eccentricity of \toib are consistent with a proto-Hot Jupiter that is undergoing high-eccentricity tidal migration and will have an orbital radius of \afinal{} au (i.e., an orbital period of \pfinal{} days) after complete tidal circularization (see \citealt{daws18} \S3.2 for a detailed discussion). \changes{The timescale for the circularization can be estimated using the equilibrium tide model \citep{hut81}, although the estimation has significant uncertainty due to the poorly understood tidal dissipation parameter $Q'_p$ and how it depends on the planet's physical and orbital properties. Neglecting stellar tidal dissipation and assuming that the planet rotates synchronously with its orbit, we adopt Equation (3) in \cite{adam06} and find a characteristic circularization timescale of 2.7 Gyr for \toib, using $Q'_p = 10^6$. We note the estimated timescale could differ by orders of magnitude for different $Q'_p$ assumptions. Moreover, the $Q'_p$ is not a constant during the planetary evolution, but has a strong dependence on the planetary radius and semi-major axis. If we instead assume a constant viscous dissipation rate $\bar{\sigma}_p$ \citep{eggl89, eggl01, hans10} and adopt Equation (2), (3), and (6) in \cite{hans10}, we find the circularization timescale to be 4 Gyr, using $\bar{\sigma}_p = 5\times10^{-6}$. Again, different assumptions for  $\bar{\sigma}_p$ could lead to very different estimation on the circularization timescale.}
\changes{\toi has an estimated age of $2.14^{+0.66}_{-0.52}$ Gyr, and is expected to leave the main sequence (MS) and begin its post-MS radial expansion in $\sim$ 0.5 Gyr according to the MIST model. Comparing the host star's lifetime to the planet's circularization timescale, it is very plausible that \toib could not get fully circularized before its host star evolves, even if it acquired its eccentricity early on. If the planet's orbit has not yet been fully circularized, the increasing stellar radiation could inflate the planetary radius and speed up the orbit circularization. After the orbital circularization, the tide raised on the star could introduce stellar dissipation and the orbital decay of the newly formed Hot Jupiter. Post-MS evolution of the host star (e.g., stellar radial expansion and mass loss) could introduce orbital instability of \toib, especially if the system has planetary or stellar companions \citep[e.g.,][]{vera13, vera17, step18}, or lead to the engulfment by the host star \citep[e.g.][]{metz12, step20}.}

\changes{In a broad context,} in Figure~\ref{fig:hem}, we present the eccentricity versus semi-major axis for all confirmed \changes{close-in planets with well-constrained eccentricities (i.e., mass $<13$ M$_{\textrm{Jup}}$, period $<365$ days) taken from the NASA Exoplanet Archive (DOI 10.26133/NEA12) as of July 26th, 2021. The grey region indicates the parameter space in which planets from a large initial semi-major axis could have undergone high-eccentricity tidal migration following the constant angular momentum tracks (i.e., $a_{\rm final} = a (1-e^2) =$ constant)}. The upper boundary of the grey region is the limit to have avoided tidal disruption earlier on the migration track. When the initial orbit at the beginning of tidal circularization is highly elliptical, the initial periapse $\approx \frac{1}{2} a_{\rm final}$ because tracks of constant angular momentum are defined by $a_{\rm final} = a (1-e^2)$ and $a (1-e^2) \approx 2 a (1-e)$ for $e \rightarrow 1$. Therefore only planets with $a_{\rm final} = 2 f_p R_p (M_\star/M_p)^{1/3}$ \changes{or greater}, where $f_p=2.7$ \citep{guil11}, could have avoided tidal disruption during periapse passages. For illustrative purposes, we assume a 1 \mj{}, 1.3 \rj{} planet orbiting a Solar-mass star and derive \changes{$a_{\rm final} = a (1-e^2) =$ 0.034 au, set as the upper boundary of the grey region.}
The lower boundary of the grey region is set by the tidal circularization timescale. Planets below the region are unlikely to get tidally circularized within their systems' lifetimes. The critical limit for $a_{\rm final}$ depends on the tidal dissipation efficiency, which is not well determined. For illustrative purposes, we adopt the value \changes{$a_{\rm final} = a (1-e^2)$ = 0.1 au, set as the lower boundary of the grey region.} \changes{For planets that are formed at large semi-major axes and have been undergoing high-eccentricity tidal migration, we would expect to observe them only in the grey region if they are decoupled from perturber.}
\changes{We also show the constant angular momentum track for planets with periapses small enough that could excite planetary $f$-mode dissipation \citep{wu18, vick19}. Above the dashed line,} the planet's periapse is close enough to the host star to excite the $f$-mode oscillation that leads to rapid orbital decay from a ``Cold" Jupiter to a ``Warm" Jupiter, so we expect an absence of highly elliptical Jupiters (i.e., $e > 0.9$) above that line. We adopt the criterion in \cite{wu18}, where $a_{\rm final} = 2 f_p R_p (M_\star/M_p)^{1/3}$ and $f_p \approx 4$, and derive $a_{\rm final} = a (1-e^2) =$ 0.05 au for a 1 \mj{}, 1.3 \rj{} planet orbiting a Solar-mass star.

\toib stands out because of its extreme eccentricity. Although many giant planets in the semi-major axis range of 0.1--1 au have non-circular orbits, only a few have eccentricities high enough to be consistent with the high-eccentricity tidal migration origin. \changes{\toib joins HD 80606b \citep{naef01, wu03} as an example of proto-Hot Jupiters with an extreme eccentricity suggesting  high-eccentricity tidal migration as one possible origin of Hot Jupiters. In Figure~\ref{fig:hem}, we label a few additional interesting systems. The $y$-axis in Figure~\ref{fig:hem} is scaled to $e^2$ to emphasize planets with non-circular orbits. Beside HD 80606b, HAT-P-2b, and Kepler-1656b that have been discussed in Section~\ref{sec:intro}, HD 17156b \citep{fisc07} could undergo the tidal migration if its tidal dissipation is efficient enough. Planets below the grey region (e.g., Kepler-419b, \citealt{daws14, jack19}; KOI-1257b, \citealt{sant14}) could periodically reach the migration track if they are coupled to external companions that cyclically excite their eccentricities to higher values.}

In Figure~\ref{fig:hem}, we also observe that nearly all the planets to the left of the dashed grey line have been fully circularized, which may indicate that tidal circularization happens very quickly at $a_{\rm final}$ $<$ 0.05 au. 
\changes{Two exceptions are TOI-269b \citep{coin21} and HD 134060b \citep{udry19}, which are both planets are small planets and could have much longer tidal evolution timescale compared to giant planets. For giant planets,} the $f$-mode dissipation could aid the migration during the early stage, e.g., from 1 au to 0.2 au with an eccentricity of 0.98 to 0.9. Although the $f$-mode line separates the population, the $f$-mode dissipation is likely not responsible for the lack of planets with $0.2 < e < 0.9$, since it only works efficiently when the planet has a very large eccentricity.

A group of Hot Jupiters on moderately elliptical orbits (i.e., $0.2 < e < 0.6$) are found in the grey region with $a_{\rm final}$ of 0.05--0.1 au, but not many on extremely elliptical orbits. Many of these moderately elliptical planets were discovered with ground-based transit surveys that are not sensitive to longer orbital periods, which might explain why we do not see these moderately eccentric Jupiters' highly eccentric counterparts earlier on the tidal migration track. However, in the \Kepler sample -- which includes longer orbital periods -- \citet{daws15} found a paucity of super-eccentric proto-Hot Jupiters inconsistent with the prediction of the high-eccentricity tidal migration origin of close-in giant planets \citep{socr12}, assuming a similar share of the \Kepler Hot Jupiter population has moderately elliptical orbits as the Hot Jupiters found in ground-based surveys. 
\changes{Recent work suggests, for planets undergoing the von Zeipel-Lidov-Kozai cycles, their extreme eccentricity spike is very short \citep[e.g.,][]{ naoz11,teys13} and their transition timescale from Cold Jupiters to Hot Jupiters is short compared to the timescale the planets spends as Cold Jupiters \citep{petr15, ande17}, making migrating planets hard to detect. A similar short transition timescale is also found in mechanisms such as secular chaos \citep[e.g.,][]{teys19} and planet-planet scattering \citep[e.g.,][]{chat08, naga08}. However, none of these studies could explain the lack of super eccentric planets ($e > 0.6$) relative to moderately eccentric planets \citep[$0.2 < e < 0.6$;][]{daws15}.}
One proposed explanation is that planets undergoing tidal migration primarily orbit more metal-rich (i.e., greater than or equal to solar metallicity) stars \citep{daws13}, so the \Kepler sample lacks both moderately eccentric Hot Jupiters and super-eccentric proto-Hot Jupiters due to the \Kepler host stars' overall lower metallicity compared to ground-based surveys \citep{daws13, guo17}. \toi has a metallicity of $[{\rm M/H}] = 0.017^{+0.057}_{-0.049}$ dex, consistent with the picture of planets undergoing tidal migration primarily orbit metal-rich stars. We have not computed a formal prediction for the number of super-eccentric Jupiters in the \TESS sample, but would have not expected many due to the shorter orbital periods expected to be detected by the \TESS mission given the observing strategy. The discovery of \toib right at the upper limit of \TESS planets' most common orbital periods suggests that proto-Hot Jupiters may not be so rare among bright, nearby stars as they are in the \Kepler sample. \changes{ A final possibility is that Hot Jupiters may begin as Warm Jupiters, instead of Cold Jupiters, following in situ formation or disk migration.}

Several dynamical processes could explain the extreme eccentricity of \toib. Generally for Jupiters on the high-eccentricity tidal migration track, the possible dynamical processes include planet-planet scattering \citep[e.g.,][]{rasi96, chat08, naga08}, stellar/planetary Kozai cycles with tidal friction \citep[e.g.,][]{wu03, naoz12, petr15b}, and secular chaos \citep[e.g.,][]{wu11}. The stellar metallicity trend \citep{daws13} supports mechanisms involving planet-planet interactions (e.g., planet-planet scattering, secular chaos, and planetary Kozai). 
\changes{The short observing baseline and limited radial velocity precision hinder the detection of radial velocity accelerations in the \toi system. From the \gaia EDR3 astrometry \citep{lind21}, the astrometric noise excess significance is 6.38, which indicates moderate astrometric excess noise. However, the renormalised unit weight error (RUWE) value of \toi is 0.929, which indicates the excess noise is unlikely to be from orbital motion. The low RUWE value and the factor that Gaia did not report any additional source within 5\arcsec of \toi can be used to place some limited constraint on mass of bound companions within 1.5--6.5 AU \citep{Penoyre:2020}, as well as beyond $\sim$ 30 AU \citep{Ziegler:2018,Rizzuto:2018}. Future Gaia astromeric limit with longer baseline will help to improve such a constraint. Long-term, precise radial-velocity observations will be key to searching for companions in the \toi system. High-resolution speckle imaging will be helpful to identify or rule out nearby binaries.
Excitation of the mutual inclination might also happen during the dynamical interactions. The planet is at wide enough separation to presumably not realigned by stellar tides. A Rossiter-McLaughlin measurement would, therefore, help to reveal the dynamical history of the planet.}

The extreme eccentricity of \toib makes it an exciting target for atmospheric observations. The incident flux received by \toib at periapse is $\sim$80 times higher than the flux it receives at apoapse. The equilibrium temperature at periapse is $\sim$ 2500\,K, comparable to classic super Hot Jupiters such as WASP-121 \citep{Delrez:2016,Evans:2017}. At apoapse, the equilibrium temperature is $\sim$ 800\,K. As the planet cools down, its atmosphere will cross the equal abundance boundaries of CO/CH$_4$ and N$_2$/NH$_3$ \citep{Fortney:2020}. Significant condensation of clouds is also expected due to the dramatic temperature change \citep{Wakeford:2017}. \toib provides an extreme case for studies of dynamic exoplanet atmosphere chemistry. The extreme time varying irradiation would also result in strong super-rotating jets in the atmosphere of \toib. When observed in relatively long wavelengths near the periapse of the orbit, we expect significant flux change \citep[i.e.][]{Laughlin:2009,Mayorga:2021} and a ``ringing" effect due to the planet's rotation \citep{Kataria:2013}. These flux variations can be easily measured by the MIRI instrument on the James Webb Space Telescope and lead to the direct detection of the rotation period of the planet.
%Previous observations made on modest eccentric planets HAT-P-2b (Lewis et al 2013), WASP-14 b (Wong et al 2015), and Kepler-434 b (Dittman et al AAS)
\toib may also experience significant tidal heating due to its high eccentricity. Adopting the tidal heating model in \citet{Leconte:2010}, we estimate the tidal energy dissipation rate is $10^{28}$\,erg\,s$^{-1}$, which is 10\% of the orbit averaged irradiation energy the planet receives from the star. Future atmosphere observations exploring the Pressure-Temperature profile of the planet may reveal the impact of such tidal heating.   

\acknowledgments
\changes{We appreciate the referee for a helpful report and in particular for comments that improve the discussion of the work. We thank Alex Venner for a helpful discussion on the interpretation of the \gaia astrometry.}

This research made use of $\mathtt{exoplanet}$ \citep{exoplanet:exoplanet, fore21} and its dependencies \citep{exoplanet:agol20, exoplanet:astropy13, exoplanet:astropy18, exoplanet:exoplanet, exoplanet:foremanmackey17, exoplanet:foremanmackey18, exoplanet:kipping13, exoplanet:luger18, exoplanet:pymc3, exoplanet:theano}. Computations for this research were performed on the Pennsylvania State University’s Institute for CyberScience Advanced CyberInfrastructure (ICS-ACI). This content is solely the responsibility of the authors and does not necessarily represent the views of the Institute for CyberScience. The Center for Exoplanets and Habitable Worlds is supported by the Pennsylvania State University and the Eberly College of Science.

This work makes use of observations from the LCOGT network. Part of the LCOGT telescope time was granted by NOIRLab through the Mid-Scale Innovations Program (MSIP). MSIP is funded by NSF. 

 This research has used data from the CTIO/SMARTS 1.5m telescope, which is operated as part of the SMARTS Consortium by RECONS (www.recons.org) members Todd Henry, Hodari James, Wei-Chun Jao, and Leonardo Paredes. At the telescope, observations were carried out by Roberto Aviles and Rodrigo Hinojosa.

Data presented herein were obtained with the MINERVA-Australis facility at the Mt. Kent Observatory from telescope time allocated through the NN-EXPLORE program. NN-EXPLORE is a scientific partnership of the National Aeronautics and Space Administration and the National Science Foundation.

The CTIO/SMARTS 1.5m and MINERVA-Australis telescope time were granted by the NOIRLab program 2021A-0147 (PI: J. Dong). 

This work has made use of data from the European Space Agency (ESA) mission {\it Gaia} (\url{https://www.cosmos.esa.int/gaia}), processed by the {\it Gaia} Data Processing and Analysis Consortium (DPAC, \url{https://www.cosmos.esa.int/web/gaia/dpac/consortium}). Funding for the DPAC has been provided by national institutions, in particular the institutions participating in the {\it Gaia} Multilateral Agreement. 

We acknowledge the use of TESS High Level Science Products (HLSP) produced by the Quick-Look Pipeline (QLP) at the TESS Science Office at MIT, which are publicly available from the Mikulski Archive for Space Telescopes (MAST). Funding for the TESS mission is provided by NASA's Science Mission directorate.

This paper includes data collected by the \TESS mission, which are publicly available from the Mikulski Archive for Space Telescopes (MAST). Resources supporting this work were provided by the NASA High-End Computing (HEC) Program through the NASA Advanced Supercomputing (NAS) Division at Ames Research Center for the production of the SPOC data products.

This research has made use of the NASA Exoplanet Archive, which is operated by the California Institute of Technology, under contract with the National Aeronautics and Space Administration under the Exoplanet Exploration Program.

This research has made use of the Exoplanet Follow-up Observation Program website, which is operated by the California Institute of Technology, under contract with the National Aeronautics and Space Administration under the Exoplanet Exploration Program.

This research received funding from the European Research Council (ERC) under the European Union's Horizon 2020 research and innovation programme (grant agreement n$^\circ$ 803193/BEBOP), and from the Science and Technology Facilities Council (STFC; grant n$^\circ$ ST/S00193X/1).

This work makes use of observations from the ASTEP telescope. ASTEP benefited from the support of the French and Italian polar agencies IPEV and PNRA in the framework of the Concordia station program and from Idex UCAJEDI (ANR-15-IDEX-01).

MINERVA-Australis is supported by Australian Research Council LIEF Grant LE160100001, Discovery Grant DP180100972, Mount Cuba Astronomical Foundation, and institutional partners University of Southern Queensland, UNSW Sydney, MIT, Nanjing University, George Mason University, University of Louisville, University of California Riverside, University of Florida, and The University of Texas at Austin. We respectfully acknowledge the traditional custodians of all lands throughout Australia, and recognise their continued cultural and spiritual connection to the land, waterways, cosmos, and community. We pay our deepest respects to all Elders, ancestors and descendants of the Giabal, Jarowair, and Kambuwal nations, upon whose lands the Minerva-Australis facility at Mt Kent is situated.

\vspace{5mm}
\facilities{\TESS, \emph{Gaia}, LCOGT, ASTEP, CTIO:1.5m CHIRON, Minerva-Australis, Exoplanet Archive}

\software{$\mathtt{ArviZ}$ \citep{Kumar2019}, $\mathtt{AstroImageJ}$ \citep{Collins:2017}, $\mathtt{astropy}$ \citep{exoplanet:astropy13, exoplanet:astropy18}, $\mathtt{celerite}$ \citep{exoplanet:foremanmackey17, exoplanet:foremanmackey18}, $\mathtt{exoplanet}$ \citep{fore21}, $\mathtt{Jupyter}$ \citep{kluy16}, $\mathtt{Matplotlib}$ \citep{hunt07, droe16}, $\mathtt{NumPy}$ \citep{vand11, harr20}, $\mathtt{pandas}$ \citep{mckinney-proc-scipy-2010}, $\mathtt{PyMC3}$ \citep{exoplanet:pymc3}, $\mathtt{SciPy}$ \citep{2020SciPy-NMeth}, $\mathtt{starry}$ \citep{exoplanet:luger18}, $\mathtt{TAPIR}$ \citep{Jensen:2013}, $\mathtt{Theano}$ \citep{exoplanet:theano}, $\mathtt{EXOFASTv2}$ \citep{Eastman:2013, Eastman:2019}}

\appendix
\section{Radial Velocities for TOI-3362}
In Table~\ref{tbl:rvs}, we present radial velocity data obtained from the \chiron and \minerva spectrographs.
\begin{table*}
\footnotesize
\setlength{\tabcolsep}{5pt}
\renewcommand{\arraystretch}{1}
\centering
\caption{Radial velocities for \toi. \label{tbl:rvs}}
\begin{tabular}{lllc}
  \hline
  \hline
BJD & RV ($\mathrm{m\,s}^{-1}$) & $\sigma_{\rm RV}$  ($\mathrm{m\,s}^{-1}$) & Instrument \\
\hline
2459319.576 & 6423.3 & 135.8 & CHIRON \\
2459320.613 & 6329.0 & 224.7 & CHIRON \\
2459321.737 & 6463.0 & 165.2 & CHIRON \\
2459322.642 & 6783.3 & 118.6 & CHIRON \\
2459323.634 & 6655.6 & 123.0 & CHIRON \\
2459324.630 & 7153.7 & 148.8 & CHIRON \\
2459327.580 & 5846.3 & 242.0 & CHIRON \\
2459329.568 & 6333.2 & 132.9 & CHIRON \\
2459330.579 & 6344.9 & 127.8 & CHIRON \\
2459332.550 & 6226.9 & 179.4 & CHIRON \\
2459333.592 & 6538.0 & 190.0 & CHIRON \\
2459334.584 & 6281.0 & 137.0 & CHIRON \\
2459335.551 & 6835.0 & 321.0 & CHIRON \\
2459342.644 & 6955.0 & 114.0 & CHIRON \\
2459343.571 & 6894.5 & 127.8 & CHIRON \\
2459343.648 & 6635.2 & 158.0 & CHIRON \\
2459344.548 & 6209.6 & 299.3 & CHIRON \\
2459345.525 & 6249.9 & 143.0 & CHIRON \\
2459345.601 & 6345.5 & 137.8 & CHIRON \\
2459346.545 & 6109.4 & 152.6 & CHIRON \\
2459346.628 & 5915.4 & 176.0 & CHIRON \\
\hline
2459351.004 & 7667.0 & 137.0 & Minerva-Australis \\
2459355.986 & 7613.0 & 170.0 & Minerva-Australis \\
2459358.939 & 7677.0 & 141.0 & Minerva-Australis \\
2459360.963 & 7928.0 & 223.0 & Minerva-Australis \\
2459361.889 & 7674.0 & 199.0 & Minerva-Australis \\
2459361.974 & 7137.0 & 250.0 & Minerva-Australis \\
2459362.885 & 7415.0 & 183.0 & Minerva-Australis \\
2459364.020 & 7391.0 & 123.0 & Minerva-Australis \\
2459364.887 & 7630.0 & 186.0 & Minerva-Australis \\
\hline
\end{tabular}
\end{table*}

\bibliography{toi3362}{}
\bibliographystyle{aasjournal}

\end{document}